\documentclass[]{iopart}
\usepackage{graphics}
\usepackage{cite}
\usepackage{amssymb}
\newcommand{\ud}{\mathrm{d}}
\begin{document}
\title{Shrinking of fluctuation region in a two-band superconductor} 
\author{Artjom Vargunin and Teet \"Ord}
\address{Institute of Physics, University of Tartu, 4 T\"ahe Street,
51010 Tartu, Estonia}


\begin{abstract} In a two-band superconductor, two qualitatively different fluctuation modes related to the gap modules contribute to free energy and heat capacity, in addition to the phase fluctuations. The first mode has divergent temperature behaviour since it accounts for the critical fluctuations around the phase transition point, $T_\mathrm{c}$, along with pseudo-critical ones associated with former instability of the weaker-superconductivity component. The involvement of these two factors, competing under interband interaction, results in the Ginzburg number which increases with $T_\mathrm{c}$ non-monotonically, allowing the reduction up to $75\%$. This makes fluctuations effective for revealing additional superconducting component in the system. The second mode does not diverge, but has a jump at $T_\mathrm{c}$, defined uniquely by the strength of interband interaction. This mode contributes fundamentally beyond critical domain.
\end{abstract}



\section{Introduction}

The behaviour of fluctuations around the critical temperature is the most principal aspect for the theory of phase transitions. Despite the phenomenological picture given by Landau, there is always a critical region where the deviations from the common description appear in the form of universality classes. In particular, the superconductivity can be treated in terms of the Ginzburg-Landau theory in striking vicinity of the phase transition temperature, however, the scaling predictions for the criticality of superconducting instability seem to follow $XY$ model \cite{XY1,XY2}, similarly to the normal-superfluid transition in the $^4$He liquid, for instance. The quantitative estimation for the temperature interval near critical point, where fluctuation cannot be considered in the Gaussian-Ginzburg-Landau approximation, is given by the Ginzburg criterion. The latter assesses the corrections stemming from the critical behaviour as unobservable in the conventional bulk superconductors.

Extraordinary properties of cuprates \cite{cuprates} have started an era of high-$T_\mathrm{c}$ superconductors thirty years ago. Nowadays, that promising family also includes magnesium diboride \cite{MgB}, fullerides \cite{fullerides}, pnictides \cite{pnictides}, to name a few. Unlike conventional superconductors, in these compounds the role of critical fluctuations becomes essential. High values of thermal disordering energy, $k_\mathrm{B}T_\mathrm{c}$, together with small coherence volume due to short correlation lengths makes it energetically inexpensive to create fluctuating patches of normal material in the superconducting state. As a result, the critical region becomes expanded by several orders of magnitude, reaching about $1$ K, or even $10$ K, depending on the system dimensionality \cite{plakida}. Additionally, strong magnetic fields, comparable to $H_{\mathrm{c}2}$, suppress the superconductivity by enhancing substantially the role of fluctuations. That enables detailed experimental study of the critical domain.

Numerous experiments indicate the significant impact of the superconducting fluctuations on various high-$T_\mathrm{c}$ systems, e.g. the reduction of transition temperature \cite{Tc} and melting of the Abrikosov flux lattice into a vortex liquid \cite{melting1,melting2}. The latter has practical consequences for the applications of high-$T_\mathrm{c}$ superconductors, since the lack of vortex pinning results in a lost of zero resistance. On the other side, preformed Cooper pairs observed above phase transition point are considered as possible explanation for various anomalies in the normal state of the underdoped copper oxides referred as pseudogap physics \cite{pseudogap}.

Although the mechanism behind high-$T_\mathrm{c}$ superconductivity is far from being clear, the contributions from more then one carrier band can play decisive role here. This picture is supported by direct experimental evidences of multicomponent scenario in a variety of relevant materials, see Refs. \cite{cuprates1,MgB1,pnictides1}, for example. At that, multigap nature opens possibilities for the fluctuation regimes not peculiar for single-gap superconductors. For instance, small interband-phase fluctuations result in collective excitations, Leggett mode \cite{leggett1,leggett2,leggett3}, which become massless and mix with density fluctuations in three-gap system at time-reversal-symmetry-breaking transition \cite{leggett4}. Phase kinks can be generated in the non-equilibrium current-carrying states \cite{gurevich}. Recently, the presence of magnetic flux-carrying metastable topological soliton was predicted in three-band superconductors with broken time-reversal symmetry \cite{babaevsoliton}.

Spatial coherency of gap order parameter plays central role in the formation of superconducting fluctuations. However, these properties become affected by multiband physics in a quite non-trivial way. Neither type-I nor type-II magnetic response, called type-1.5 behaviour \cite{babaev0,moshchalkov0,ray}, is such an example. The phenomenon is based on the existence of several competing coherency channels for inhomogeneous multiband superconductor. Due to interband pairing, the corresponding correlation lengths describe the joint superconducting condensate as a whole, not definite bands involved \cite{babaev1,babaev1a,babaev1b,ord1,vargunin1,litak1,litak2}. These characteristic length scales naturally appear in the spatial behaviour of superconductivity fluctuations and in the spatial correlation functions.

In this paper, we investigate the spatially inhomogeneous thermal fluctuations of superconductivity in the case when competing coherency scales are present. We analyse the interplay between fluctuation modes related to these length scales. The main attention was previously centered only upon contribution of critical fluctuations to specific heat and conductivity in multiband scenario \cite{koshelev,fanfarillo,askerzade}. However, the competition of superconductivity components was never clarified in this respect. Here we also give special attention to the coherency channel which is characterized by correlation length with non-diverging temperature behaviour. As a result, we find a way to reveal an additional superconducting component in the system by inspecting fluctuations around phase transition point. The findings may be also of relevance for elementary superconductors, since they become effectively multiband in the nanometre scale \cite{nano,nano1}.

\section{Superconductivity fluctuations}We start with two-component ($\alpha=1,\ 2$) Ginzburg-Landau functional \cite{zhitomirsky} for complex gap-order parameters  $\delta_\alpha=\Re\delta_\alpha+i\Im\delta_\alpha$ in the absence of magnetic field 
\begin{equation}\label{e1}
F=F_\mathrm{n}+\sum_\alpha\int
\Big(
a_\alpha|\delta_\alpha|^2+\frac{b_\alpha}{2}|\delta_\alpha|^4-c\delta_\alpha\delta_{3-\alpha}^\ast+K_\alpha|\nabla\delta_\alpha|^2\Big)\ud
V,
\end{equation}
where $F_\mathrm{n}$ is free energy without superconductivity. For expansion coefficients we retain full temperature dependence, i.e.
\begin{eqnarray}\label{e2}
&&a_\alpha=\frac{W_{3-\alpha,3-\alpha}}{W^2}-\rho_\alpha\ln\frac{1.13\hbar\omega_\mathrm{D}}{k_\mathrm{B}T},\qquad b_\alpha=\frac{0.11\rho_\alpha}{(k_\mathrm{B}T)^2},\\
&&c=\frac{W_{12}}{W^2},\qquad K_\alpha=\frac{0.02\rho_\alpha\hbar^2v_{\mathrm{F}\alpha}^2}{(k_\mathrm{B}T)^2},\qquad W^2=W_{11}W_{22}-W_{12}^2. \nonumber
\end{eqnarray}
Here $W_{\alpha\alpha}>0$ and $W_{12}=W_{21}$ are matrix elements for intraband and interband pair-transfer interaction channels, $\rho_\alpha$ is the density of states at the Fermi level, and $v_{\mathrm{F}\alpha}$ is the Fermi
velocity in the corresponding band. Electron-electron interactions are assumed to be non-zero and independent on electron wave
vector in the Debye layer $\pm\hbar\omega_\mathrm{D}$ around chemical potential.

In the homogeneous case the minimization of Ginzburg-Landau functional
leaves us with equations for coupled bulk order parameters $\Delta_\alpha=|\Delta_\alpha|e^{i\phi_\alpha}$, namely,
\begin{equation}\label{e3}
\Delta_\alpha(a_\alpha+b_\alpha|\Delta_\alpha|^2)=c\Delta_{3-\alpha},\quad\mathrm{sqn}\big(c\cos(\phi_1-\phi_2)\big)=1,
\end{equation}
where $\mathrm{sqn}$ is sign function. These equations fix the modules of the order parameters and the difference between the phases $\phi_1-\phi_2=\pi n$, $n\in N$. Since the gap phases remain unspecified we take real bulk order parameters $\Delta_\alpha=\Delta_\alpha^\ast$.

The condition for the critical point
$a_{1\mathrm{c}}a_{2\mathrm{c}}=c^2$ (here and elsewhere index "c" implies
$T=T_\mathrm{c}$) has two solutions
$T_{\mathrm{c}\pm}$ which transform into the separate intraband transition
temperatures $T_{\mathrm{c}\alpha}\sim \exp(\frac{-1}{\rho_\alpha W_{\alpha\alpha}})$ in the limit $W_{12}\to0$. The larger
solution $T_{\mathrm{c}-}$ corresponds to the superconducting phase transition temperature of the
joint condensate, denoted as $T_\mathrm{c}$, and it increases with
interband coupling constant $|W_{12}|$. The smaller solution $T_{\mathrm{c}+}$ is a monotonically decreasing function of $|W_{12}|$ which disappears as $W\to0$.
These dependencies are explicitly depicted in Ref. \cite{vargunin5}.

\subsection{Free energy fluctuations}
In the macroscopic system, homogeneous state with free energy $F_\mathrm{h}$ is very
probable and fluctuation effects are bound to the inhomogeneity of the gap order parameters. To describe statistically small deviations from the homogeneous superconducting ($\Delta_{\alpha}\neq 0$) or normal ($\Delta_{\alpha}= 0$) background, we take $\delta_\alpha(\mathbf{r})=\Delta_\alpha+\eta_\alpha(\mathbf{r})$ and use the Gaussian approximation. In terms of complex Fourier components
$\Re\delta_{\alpha\mathbf{k}}=\frac{1}{V}\int\Re\delta_\alpha(\mathbf{r})e^{-i\mathbf{kr}}\ud V=\Delta_\alpha \delta_{\mathbf{k},0}+\Re\eta_{\alpha\mathbf{k}}$ (here $\delta_{\mathbf{k},0}$ is Kronecker delta) and $\Im\delta_{\alpha\mathbf{k}}=\frac{1}{V}\int\Im\delta_\alpha(\mathbf{r})e^{-i\mathbf{kr}}\ud V=\Im\eta_{\alpha\mathbf{k}}$, the Ginzburg-Landau functional reads as
\begin{eqnarray}\label{e4}
&&F=F_\mathrm{h}+V\sum_\alpha\Big[A_\alpha\Re\eta_{\alpha\mathbf{0}}^2+B_\alpha\Im\eta_{\alpha\mathbf{0}}^2-c\big(\Re\eta_{\alpha\mathbf{0}}\Re\eta_{3-\alpha\mathbf{0}}+\Im\eta_{\alpha\mathbf{0}}\Im\eta_{3-\alpha\mathbf{0}}\big)+\nonumber\\
&&2\sum_{|\mathbf{k}|\neq0}^\star\Big(A_{\alpha\mathbf{k}}|\Re\eta_{\alpha\mathbf{k}}|^2+B_{\alpha\mathbf{k}}|\Im\eta_{\alpha\mathbf{k}}|^2-c\big(\Re\eta_{\alpha\mathbf{k}}\Re\eta_{3-\alpha\mathbf{k}}^\ast+\Im\eta_{\alpha\mathbf{k}}\Im\eta_{3-\alpha\mathbf{k}}^\ast\big)\Big)\Big],\nonumber\\
\end{eqnarray}
where $A_{\alpha\mathbf{k}}=A_\alpha+K_\alpha\mathbf{k}^2$,
$A_\alpha=a_\alpha+3b_\alpha\Delta_\alpha^2\geq0$ and $B_{\alpha\mathbf{k}}=B_\alpha+K_\alpha\mathbf{k}^2$,
$B_\alpha=a_\alpha+b_\alpha\Delta_\alpha^2\geq0$. The star sign near summation denotes the half of $\mathbf{k}$-space. In this presentation $F$ contains only independent degrees of freedom: real $\Re\eta_{\alpha\mathbf{0}}$, $\Im\eta_{\alpha\mathbf{0}}$ and real and imaginary parts of $\Re\eta_{\alpha\mathbf{k}}$, $\Im\eta_{\alpha\mathbf{k}}$ taken in half of $\mathbf{k}$-space. To calculate statistical sum $Z$ one should integrate $\exp\big(-\frac{F}{k_\mathrm{B}T}\big)$ over these variables. Note that interband pairing results in the non-diagonal elements in quadratic form (\ref{e4}), i.e. some degrees of freedom become mixed. As a result, for macroscopic superconductor one obtains free energy density $f=-\frac{k_\mathrm{B}T}{V}\ln Z$ in the form
\begin{eqnarray}\label{e5}
&&f=\frac{F_\mathrm{h}}{V}+\frac{k_\mathrm{B}T}{V}\sum_{k\neq0}^\star\ln
\big[(k^2+\kappa_+^2)(k^2+\kappa_-^2)(k^2+m_+^2)(k^2+m_-^2)\big],\nonumber\\
&&\kappa_\pm^2=\frac{1}{2}\sum\limits_\alpha\xi_\alpha^{-2}\pm\sqrt{\Big(\frac{\xi_1^{-2}-\xi_2^{-2}}{2}\Big)^2+\frac{c^2}{K_1K_2}},\\
&&m_\pm^2=\frac{1}{2}\sum\limits_\alpha m_\alpha^2\pm\sqrt{\Big(\frac{m_1^2-m_2^2}{2}\Big)^2+\frac{c^2}{K_1K_2}},\nonumber
\end{eqnarray}
with $\xi_\alpha^2=\frac{K_\alpha}{A_\alpha}$ and $ m_\alpha^2=\frac{B_\alpha}{K_\alpha}$. For the mass factors one obtains $m_\pm^2=\kappa_\pm^2$ above $T_\mathrm{c}$ and $m_+^2=\sum_\alpha m_\alpha^2$, $m_-=0$ below $T_\mathrm{c}$. These formulas generalize
single-band ($W_{12}\to0$) consideration for which gap-amplitude fluctuations contribute as $\frac{k_\mathrm{B}T}{V}\sum\limits_{k\neq0}^\star\ln(k^2+\xi^{-2})$
and gap-phase fluctuations as $\frac{k_\mathrm{B}T}{V}\sum\limits_{k\neq0}^\star\ln(k^2+m^2)$, see Ref. \cite{larkin}. Here $\xi$ is the correlation length of the one-gap system diverging at
$T_\mathrm{c}$ and $m$ is the mass of Goldstone boson disappearing below $T_\mathrm{c}$. In a two-band superconductor there appear four
fluctuation channels related to the distinct coherency lengths $\kappa_\pm^{-1}$ and to the mass factors associated with Goldstone mode ($m_-$) and Leggett mode ($m_+$).

Various experimental techniques, e.g. scanning tunneling spectroscopy, muon spin relaxation and thermal conductivity measurements, point to the evidences of distinct coherency scales for two-band superconductivity \cite{exp1,exp2,exp3}. In our approach the correlation lengths have non-critical ($\kappa_+^{-1}$) and critical ($\kappa_-^{-1}$) temperature behaviour near superconducting phase transition point as follows from the expansions
\begin{eqnarray}\label{e6}
&&\kappa_-^2\approx(\rho_-+\Delta\rho_-)\frac{T-T_\mathrm{c}}{T_\mathrm{c}},\quad\rho_-=\frac{\frac{\rho_1}{a_{1\mathrm{c}}}+\frac{\rho_2}{a_{2\mathrm{c}}}}{\xi_{1\mathrm{c}}^2+\xi_{2\mathrm{c}}^2},\nonumber\\
&&\kappa_+^2\approx\xi_{1\mathrm{c}}^{-2}+\xi_{2\mathrm{c}}^{-2}+\left(\rho_++\Delta\rho_+\right)\frac{T-T_\mathrm{c}}{T_\mathrm{c}},\\
&&\rho_+=2\kappa_{+\mathrm{c}}^2+\frac{\rho_-}{\xi_{1\mathrm{c}}^2\xi_{2\mathrm{c}}^2}\frac{\frac{\rho_1}{a_{1\mathrm{c}}}\xi_{2\mathrm{c}}^4+\frac{\rho_2}{a_{2\mathrm{c}}}\xi_{1\mathrm{c}}^4}{\frac{\rho_1}{a_{1\mathrm{c}}}+\frac{\rho_2}{a_{2\mathrm{c}}}},\nonumber
\end{eqnarray}
where $\Delta\rho_\pm=0$ above $T_\mathrm{c}$, however, below $T_\mathrm{c}$ one
should take $\Delta\rho_-=-3\rho_-$ and
\begin{equation}\label{e7}
\Delta\rho_+=-\frac{3\rho_-}{\xi_{1\mathrm{c}}^2\xi_{2\mathrm{c}}^2}\frac{\frac{\rho_1}{a_{1\mathrm{c}}^2}\xi_{2\mathrm{c}}^4+\frac{\rho_2}{a_{2\mathrm{c}}^2}\xi_{1\mathrm{c}}^4}{\frac{\rho_1}{a_{1\mathrm{c}}^2}+\frac{\rho_2}{a_{2\mathrm{c}}^2}}.\nonumber
\end{equation}
Note that same length scales $\kappa_\pm^{-1}$ follow also directly from the Ginzburg-Landau gap equations \cite{vargunin4}.

For the Leggett mode we have obtained the same expression as given in Ref. \cite{babaev1a}. For this phase difference mode we get
\begin{eqnarray}\label{e8}
&&m_+^2\approx\xi_{1\mathrm{c}}^{-2}+\xi_{2\mathrm{c}}^{-2}+(\rho_++\Delta m_+)\frac{T-T_\mathrm{c}}{T_\mathrm{c}},
\end{eqnarray}
where $\Delta m_+=0$ above $T_\mathrm{c}$, but below $T_\mathrm{c}$ one
should take $\Delta m_+=\frac{\Delta\rho_+}{3}$. This mass parameter describes the deviations from the equilibrium value of interband phase.

\subsection{Heat capacity fluctuations}
Leading contribution to fluctuation specific heat and conductivity, stemming from critical behaviour of correlation length, was previously considered in Refs. \cite{koshelev,fanfarillo,askerzade}. Next we calculate specific heat by taking into account all four modes. By using integration in half-space instead of
summation $\sum_\mathbf{k}\to\frac{V_d}{(2\pi)^d}\int\ud\mathbf{k}$, one obtains
specific heat capacity $c=-Tf^{\prime\prime}$ in the form
$c=c_\mathrm{h}+\sum_\tau\mathcal{F}_d(\tau)$, where
$c_\mathrm{h}$ is related to the bulk homogeneous state and
\begin{eqnarray}\label{e9}
&&\mathcal{F}_d(\tau)=\Bigg\{\begin{array}{ccc}\frac{k_\mathrm{B}T}{(2\pi)^2}\Big(\frac{2}{3}\theta_3(\tau)\tan^{-1}\frac{k_\mathrm{max}}{\tau}-\theta_2(\tau)
k_\mathrm{max}-\frac{T}{2}\frac{k_\mathrm{max}\tau^{2\prime2}}{k_\mathrm{max}^2+\tau^2}\Big),&\\\frac{k_\mathrm{B}TS}{8\pi
V}\Big(\frac{\tau^{2\prime2}}{\tau^2}\frac{Tk_\mathrm{max}^2}{k_\mathrm{max}^2+\tau^2}-\theta_2(\tau)\ln\big(\frac{k_\mathrm{max}^2}{\tau^2}+1\big)\Big),&\\\frac{k_\mathrm{B}TL}{\pi
V}\Big(\frac{Tk_\mathrm{max}\tau^{\prime2}}{k_\mathrm{max}^2+\tau^2}-\theta_1(\tau)\tan^{-1}\frac{k_\mathrm{max}}{\tau}
\Big),&\end{array}
\end{eqnarray}
for different effective dimensionality $d=3, 2, 1$ ($V_d=V, S, L$),
correspondingly. Here $\tau$ takes values $\kappa_+,\kappa_-,m_+,m_-$, prime implies temperature derivative, $\theta_n(\tau)=2\tau^{n\prime}+T\tau^{n\prime\prime}$ and
$k_\mathrm{max}$ is the cut-off parameter. Note that massless Goldstone mode does not contribute to the heat capacity fluctuations in superconducting state and in the normal phase one has $\sum_\tau\mathcal{F}_d(\tau)=2(\mathcal{F}_d(\kappa_+)+\mathcal{F}_d(\kappa_-))$.  Due to the critical behaviour of $\kappa_-^{-1}$, the most
dominating contribution from superconductivity fluctuations in the normal phase
near $T_\mathrm{c}$ has a standard form
\begin{eqnarray}\label{e10}
&&\sum_\tau\mathcal{F}_d(\tau)\approx k_\mathrm{B}\vartheta_d\rho_-^\frac{d}{2}\frac{V_d}{
V}\Big(\frac{T-T_\mathrm{c}}{T_\mathrm{c}}\Big)^{\frac{d}{2}-2},
\end{eqnarray}
where $\vartheta_d=2^{-d}\pi^{-\frac{d}{2}}\Gamma(2-\frac{d}{2})$. Symmetrically below $T_\mathrm{c}$ fluctuation contribution differs from value (\ref{e10}) by factor $2^{\frac{d}{2}-1}$. At the same time,
$\mathcal{F}_d(\kappa_+)$ stays finite by approaching critical point. 

\subsection{Ginzburg number}
The way in which the Ginzburg number can be introduced is not unique \cite{larkin}. The estimate of the critical region can be found by comparing the fluctuation energy with superconducting condensation energy, or by comparing the Aslamazov-Larkin correction to conductivity with its normal value. Here we use an approach based on the corrections to heat capacity. Note that corresponding definitions of the Ginzburg number differ by the numeric factors only.

By evaluating gaps in the homogeneous state and heat capacity jump at
$T_\mathrm{c}$, namely
\begin{equation}\label{e11}
\Delta c=c_\mathrm{hc}-c_\mathrm{nc}=\frac{1}{T_\mathrm{c}}\frac{\big(\frac{\rho_1}{a_{1\mathrm{c}}}+\frac{\rho_2}{a_{2\mathrm{c}}}\big)^2}{\frac{b_{1\mathrm{c}}}{a_{1\mathrm{c}}^2}+\frac{b_{2\mathrm{c}}}{a_{2\mathrm{c}}^2}},
\end{equation}
one estimates the Ginzburg number by comparing the fluctuation contribution
(\ref{e10}) and $\Delta c$. The width for the temperature region affected
enormously by the fluctuations reads as
\begin{equation}\label{e12}
Gi_d=\Bigg(k_\mathrm{B}\vartheta_d\frac{\rho_-^\frac{d}{2}}{\Delta c}\frac{V_d}{
V}\Bigg)^\frac{2}{4-d}.
\end{equation}
The latter expression is analogous to the single-band counterpart. However, here
$\rho_-$ and $\Delta c$ are generally no longer power-law function of the critical temperature as for usual superconductivity. As a result, one-band power-low scaling $Gi_d\sim T_\mathrm{c}^\frac{2d-2}{4-d}$ does not hold, and
$Gi_d(T_\mathrm{c})$ becomes more general function in two-gap case.

\section{Discussions}
\begin{figure}
\begin{center}
\resizebox{0.7\columnwidth}{!}{\includegraphics{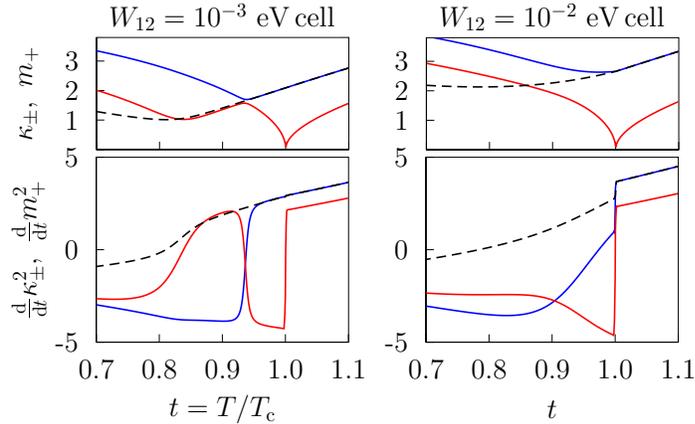}}
\end{center}
\vspace{-0.2cm}\caption{Above: The plots of inverse correlation lengths
$\kappa_-$ (red), $\kappa_+$ (blue) and mass of the Leggett mode $m_+$ (black) in units $\frac{10^{-2}}{\mathrm{nm}}$ \textit{vs} temperature for different
interband couplings. Below: the temperature derivative of $\kappa_-^2$ (red), $\kappa_+^2$ (blue) and $m_+^2$ (black) in units $\frac{10^{-3}}{\mathrm{nm}^2}$ on the same scale. The model parameters are
$W_{11,22}=0.3\mathrm{\  eV\,cell}$, $\rho_{1,2}=(1,0.94)\mathrm{\
(eV\,cell)}^{-1}$, $v_{\mathrm{F}1,2}=(5,5.104)\cdot10^5\mathrm{\ m/s}$,
$\textnormal{cell}=0.1\mathrm{\ nm}^3$. For these values
$T_{\mathrm{c}2}=0.81T_{\mathrm{c}1}$, and electron spectrum supposed to be parabolic
$\frac{\rho_2}{\rho_1}=\big(\frac{v_{\mathrm{F}1}}{v_{\mathrm{F}2}}\big)^3$.}\label{f1}
\end{figure}
\subsection{Inverse correlation lengths and masses}
The interaction between superconductivity components changes the properties of
the joint condensate in a remarkable manner. Without interband pairing the
condensate splits into two non-interacting superconducting subsystems with
corresponding phase transition points at $T_{\mathrm{c}\alpha}$. The latter temperatures determine the critical behaviour for the quantities related to these subsystems, e. g. divergences of correlation length and relaxation time, or jump of heat capacity. Since weak interband coupling acts on the weaker-superconductivity constituent as an external field\cite{vargunin2}, "applied" interband interaction smears and eliminates the criticality at lower $T_{\mathrm{c}\alpha}$, as well as mixes both superconductivity components.   This results, in particular, in the appearance of the correlation lengths $\kappa_\pm^{-1}$ with qualitatively different peculiarities\cite{babaev1,babaev1a,babaev1b,ord1,vargunin1}, see Fig. \ref{f1}.
Non-critical length $\kappa_+^{-1}$ changes with temperature weakly. However,
the behaviour of the critical one, $\kappa_-^{-1}$, points clearly to the phase
transition temperature of the two-gap system by diverging at $T_\mathrm{c}$. At
that, the former autonomous phase transition of the weaker-superconductivity component, smeared by the
interband coupling, becomes visible in case of tiny interband pairings as
noticeable non-monotonicity below $T_\mathrm{c}$. The latter is accompanied by
the presence of avoided crossing point between $T_{\mathrm{c}1}$ and
$T_{\mathrm{c}2}$.

In the regime of very weak interband pairings (roughly, $W_{12}^{2}\lesssim 10^{-4}W_{11}W_{22}$), the presence of avoided crossing point and the memory of weaker-superconductivity criticality play decisive role for the differential properties of correlation lengths. The first feature manifests itself in the  step-like peculiarity between $T_{\mathrm{c}1}$ and $T_{\mathrm{c}2}$ which is seen in Fig. \ref{f1} for both characteristics $\kappa_\pm^{2\prime}$ simultaneously. The second feature forces $\kappa_-^{2\prime}$ to increase rapidly  near lower $T_{\mathrm{c}\alpha}$. This behaviour reflects a jump at the phase transition point of the weaker-superconductivity constituent smeared by "external field" with intensity $W_{12}$. The increase of interband interaction suppresses the effects of both features.

The temperature derivative of $\kappa_-^2$ has a step at $T_\mathrm{c}$ which
follows "the law of 2", i. e.
$\big|\frac{\rho_-+\Delta\rho_-}{\rho_-}\big|=2$, see Fig. \ref{f1}. Due to
non-critical character, that "law" is violated for $\kappa_+^2$, however, the
mixing of the superconductivity components results in the jump of $\kappa_+^2$
derivative at $T_\mathrm{c}$ as well. The height of the jump is determined by
$\Delta\rho_+$, and for small values of $W_{12}$ one has $\Delta\rho_+\sim
W_{12}^2$. The jump becomes more pronounced in the superconductors with
stronger interband pairings, since $|\Delta\rho_+|$ increases monotonically by raising $|W_{12}|$ and approaches quickly $|\Delta\rho_-|$. Note that there is a limiting value $W\approx0$ at which real non-critical correlation length disappears from the Ginzburg-Landau approach as well as microscopic theory\cite{babaev1,babaev1a,babaev1b}. 

The mass parameter $m_+$ which reflects the excitation of the Leggett mode behaves  also non-critically. Similarly to $\kappa_+$, its temperature derivative is characterized by the finite jump $\frac{\Delta\rho_+}{3}$ at phase transition point. However, as interband coupling decreases, mass $m_+$ softens near former critical point of the weaker-superconductivity component. In the limit $W_{12}\to0$ relevant mode becomes massless at lower $T_{\mathrm{c}\alpha}$ due to Goldstone theorem.

The memory of the former criticality for the weaker-superconductivity, avoided crossing point and jump $\Delta\rho_+$ play crucial role for the heat-capacity fluctuations, since the latter are defined by derivatives of $\kappa_\pm$ and $m_+$. The contributions from the peculiarities indicated will alternate with increase of interband interaction. Although the coupling between gap order parameters is not an easily tunable parameter for experimentalist, it is still possible to vary it by changing the proximization between distinct electron subsystems \cite{proximization1,proximization2,proximization3}, e.g. through the doping, pressure, or direct adjustment.

\subsection{Heat capacity fluctuations}
\begin{figure}
\begin{center}
\resizebox{0.7\columnwidth}{!}{\includegraphics{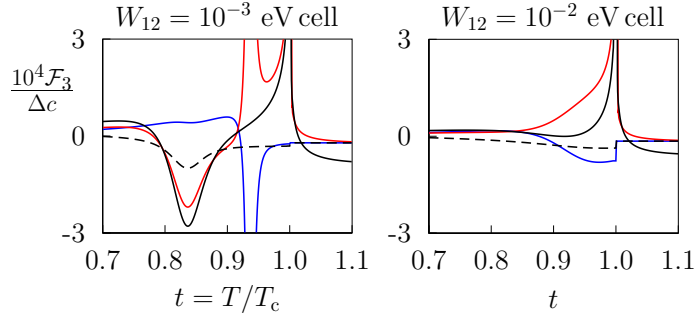}}
\end{center}
\vspace{-0.2cm}\caption{The plots of $\mathcal{F}_3(\kappa_-)$ (red), $\mathcal{F}_3(\kappa_+)$
(blue), $\mathcal{F}_3(m_+)$ (dashed black) and $\sum_\tau\mathcal{F}_3(\tau)$ (solid black) \textit{vs} temperature for different interband couplings. The model parameters are given in Fig.
\ref{f1} and $k_\mathrm{max}=0.05$ nm$^{-1}$.}\label{f2}
\end{figure}

In what follows we discuss the superconductivity fluctuations in two-band
approach. The presence of two correlation lengths and two masses allows one to identify relevant channels of fluctuation, $\mathcal{F}_d(\tau)$, contributing heat capacity. Their temperature dependencies are essentially different, see Fig. \ref{f2} for $d=3$ (the dependencies remain the same for $d=2,1$ with $y$-scale factor about $10^2,1$, correspondingly). The contribution $\mathcal{F}_d(\kappa_-)$ is strongly dominating at the phase transition point with expected divergent behaviour (\ref{e10}). Moreover, it reflects the memory of the weaker-superconductivity criticality by demonstrating a maximum with the well in the middle. This structure appears in the vicinity of $t\approx0.83$ in the left panel of Fig. \ref{f2}. The well is stemming from the differentiability of $\kappa_-^{-1}$ near $T_{\mathrm{c}2}$ instead of singularity. At that, by decreasing interband interaction, the depth of the well and the height of its borders will gradually raise together with decrease in the width of the well. For the vanishing value of $W_{12}$ the well near $T_{\mathrm{c}2}$ disappears, leaving us with well-known divergent behaviour at the lower phase transition point.

In the systems with tiny interband pairings, the avoided crossing point in the behaviours of $\kappa_\pm$ manifests itself as the extremes of $\mathcal{F}_d(\kappa_\pm)$ which compensate each other (in the left panel of Fig. \ref{f2} at $t\approx0.93$). Therefore, as the net effect from avoided crossing point is zero in a superconductor with weakly interacting gaps, both contributions $\mathcal{F}_d(\kappa_\pm)$ are equally important. By decreasing interband interaction, the extrema related to avoided crossing point raise, tighten and finally disappear for $W_{12}=0$. Note that in this limiting case $\mathcal{F}_d(\kappa_\pm)$ will be replaced by band contributions for $\alpha=1, 2$ with autonomous divergent behaviours at $T_{\mathrm{c}1}$ or $T_{\mathrm{c}2}$, correspondingly.

Interband phase fluctuations are also affected by the memory of the weaker-superconductivity criticality. Due to softening of the Leggett mode in the vicinity of the lower $T_{\mathrm{c}\alpha}$, fluctuation channel  $\mathcal{F}_d(m_+)$ contributes considerably to the heat capacity in this temperature region when interband pairing is very weak.

Whereas $\mathcal{F}_d(\kappa_+)$ and $\mathcal{F}_d(m_+)$ behave non-critically, these fluctuation modes become always overshadowed by $\mathcal{F}_d(\kappa_-)$ near $T_\mathrm{c}$. However, for sufficiently strong interband couplings, the considerable jump $\Delta\rho_+$ leads to the observable step-like temperature behaviour of $\mathcal{F}_d(\kappa_+)$ and $\mathcal{F}_d(m_+)$. As a result, there appears the discrepancy between two-band consideration, $\sum_\tau\mathcal{F}_3(\tau)$, and single-mode approach, where only unique critical mode $\mathcal{F}_3(\kappa_-)$ is taken into account. The effect is observable in the right panel of Fig. \ref{f2}, and it is sensitive to the value $k_\mathrm{max}$ chosen.

To summarize, beyond the critical region, the manifestation of fluctuations in heat capacity is qualitatively distinguishable in  two-band superconductivity model and single-gap/single-mode scenarios as follows

(i) The single-gap approach fails in the reproducing the memory of the weaker-superconductivity criticality. Intuitively, the latter effect can be incorporated by taking into consideration additional single-band subsystem, resulting in the enhancement of the fluctuations in the relevant temperature region. However, this attempt fails because a self-consistent two-gap model predicts a maximum for the fluctuations near lower $T_{\mathrm{c}\alpha}$ disguised by a deep well in the middle. The reduction of the heat capacity value near the former critical point of the weaker-superconductivity component due to fluctuations is an essential feature of a two-band scenario.

(ii) In single-mode approach there appears an enhancement of the fluctuations below phase transition point, if interband pairing is tiny (at $t\approx0.93$ in the left panel of Fig. \ref{f2}). This behaviour has no physical meaning. By increasing interband interaction constant the enhancement can be suppressed. Nonetheless, the peak of heat-capacity fluctuations remains significantly overestimated in width. These shortcomings become removed by inclusion of fluctuation mode associated with non-critical correlation length.

Note also that restricting ourselves to considering uniform spatial mode only (i.e. homogeneous system), thermal fluctuations in a finite-size two-band superconductor reveal already the memory of the former autonomous phase transition in the band with weaker superconductivity \cite{vargunin3}.

\begin{figure}
\begin{center}
\resizebox{0.7\columnwidth}{!}{\includegraphics{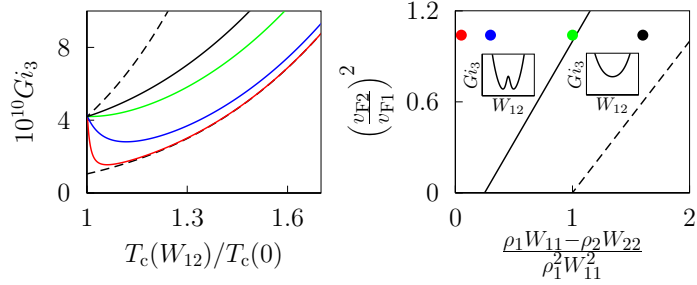}}
\end{center}
\vspace{-0.2cm}\caption{Left: The plot of $Gi_3$ \textit{vs} $T_\mathrm{c}(W_{12})$ for the same model parameters as in Fig. \ref{f1} except $W_{22}=0.31$ (red), $0.29$ (blue), $0.22$ (green) and $0.17\mathrm{\  eV\,cell}$ (black curve). For these values $\frac{T_\mathrm{c2}}{T_\mathrm{c1}}=0.95, 0.72, 0.24, 0.05$, correspondingly. The lower dashed curve represents the limiting case $W_{22}=0.32\mathrm{\  eV\,cell}$ ($T_\mathrm{c2}=T_\mathrm{c1}$). The upper dashed curve corresponds to the single-band $W_{12}\to0$ dependence on the critical temperature normalized to $T_\mathrm{c}(W_{12}=0)=T_{\mathrm{c}1}$. Although the latter temperature in our model calculations equals $23.5$ K resulting very small $Gi_3$ value, the Ginzburg number becomes magnified by several orders of magnitude by taking the parameters of a real high-$T_\mathrm{c}$ material. Right: Two regions in the parameters space where $Gi_d$ behaves monotonically or non-monotonically with $|W_{12}|$ increase. Solid line is the separation border for $d=3$, dashed line for $d=2$. Red/blue/green/black points represent the curves with the same colour as in the left panel. Note that the maximum of $Gi_{2,3}(W_{12})$ tightens as $\rho_2W_{22}$ moves to $\rho_1W_{11}$ and disappears in the limit $\rho_2W_{22}=\rho_1W_{11}$.}\label{f3}
\end{figure}
\subsection{Width of fluctuation region}
The estimations of the critical region require accurate measurements of conductivity, thermal expansion, specific heat or magnetization. Usually, the width of the critical domain is being investigated in various high-$T_\mathrm{c}$ materials only under applied magnetic fields \cite{gi1,gi2,gi3,gi4,gi5}. Next we discuss the qualitative behaviour of $Gi_d$ as a function of the critical temperature in two-band model. Fig. \ref{f3} demonstrates that the dependence can be very dissimilar to the single-band analogue for which Ginzburg number is always monotone function, $Gi_d\sim T_\mathrm{c}^\frac{2d-2}{4-d}$. Surprisingly, the presence of another superconducting component can result in the non-monotone behaviour with $T_\mathrm{c}$. 

To analyse the functional relation $Gi_d(T_\mathrm{c})$, where critical temperature changes under interband interaction, we notice that this function always follows qualitatively the dependence $Gi_d(|W_{12}|)$. At that, the Ginzburg number always increases with interband coupling, if $|W_{12}|$ exceeds some finite value. However, there are two ways how Ginzburg number changes near $W_{12}=0$. It can have global minimum or alternatively local maximum at $W_{12}=0$ (see insets in the right panel in Fig. \ref{f3}). These two regimes are separated by the condition
\begin{equation}\label{e13}
1+\frac{d}{4-d}\frac{v_{\mathrm{F}2}^2}{v_{\mathrm{F}1}^2}\gtrless\frac{2(d-1)}{4-d}\frac{\rho_1W_{11}-\rho_2W_{22}}{\rho_1^2W_{11}^2},
\end{equation}
where upper (lower) sign corresponds to the maximum (minimum) of $Gi_d$ near $W_{12}=0$. For $d=3, 2$ the separation lines in the parameters space are depicted in right panel in Fig. \ref{f3}. Formally, there should be always a local maximum at $W_{12}=0$ for $Gi_{1, 0}$. Thus, by moving from single-band to two-gap description (i.e. by turning interband interaction on), the fluctuation effects may be enhanced or suppressed as one can observe in the insets of Fig. \ref{f3}. 

The small discrepancy between intraband critical temperatures $T_{\mathrm{c}\alpha}$ and strong memory of the former instability of weaker-superconductivity component support the shrinking of the fluctuation region.
Its reduction can be enhanced by bringing together intraband critical points, $\rho_2W_{22}\to\rho_1W_{11}$ ($T_{\mathrm{c}2}\to T_{\mathrm{c}1}$), see Fig. \ref{f3}. In this process the heat-capacity jump $\Delta c$, appearing in the denominator of $Gi_d$, grows for any finite $W_{12}$. That jump increases abruptly for $W_{12}=0$, by becoming a superposition of two jumps related to superconducting subsystems involved.
As a result, one obtains effective single-band dependence $Gi_d\sim T_\mathrm{c}^\frac{2d-2}{4-d}$, when $T_{\mathrm{c}1}$ and $T_{\mathrm{c}2}$ coincide. The maximal drop of $Gi_d(T_\mathrm{c})$ can be found from the ratio
\begin{equation}\label{e14}
\frac{Gi_d(W_{12}\to0;\rho_2W_{22}\to\rho_1W_{11})}{Gi_d(W_{12}=0,\rho_2W_{22}<\rho_1W_{11})}=\frac{\left(1+\frac{\rho_1}{\rho_2}\right)^\frac{2}{4-d}}{2\left(1+\frac{v_{\mathrm{F}2}^2}{v_{\mathrm{F}1}^2}\right)^\frac{d}{4-d}}.
\end{equation}
For parabolic spectrum the latter function has a global minimum at $\frac{\rho_1}{\rho_2}=1$ ($d=3$) with the value $0.25$. For $d=2,1$ that minimum is higher. Thus, $Gi_d$ can be reduced up to $75\%$ by increasing $T_\mathrm{c}$. 

We notice that in the case $T_{\mathrm{c}2}=T_{\mathrm{c}1}$ the Ginzburg number behaves actually non-monotonically in the vicinity of $W_{12}=0$. This is caused by the non-critical correlation length which becomes divergent $\kappa_{+\mathrm{c}}^2=0$ for $W_{12}=0$. As a result, corresponding fluctuation channel $\mathcal{F}_d(\kappa_+)$ contributes to the critical region for $W_{12}=0$ increasing its width. Consequently, the non-monotonicity of $Gi_d(T_\mathrm{c})$ is related to the interplay between criticalities of superconductivity components driven by interband pairing. 

The interpretation of the $Gi_d(T_\mathrm{c})$ behaviour in two-band system can be given as follows. Let us fix intraband critical temperatures $T_{\mathrm{c}\alpha}$ by fixing parameters $\rho_\alpha$ and $W_{\alpha\alpha}$, $\alpha=1,2$. By turning interband interaction on, these two points become replaced by the phase transition temperature of the joint condensate, $T_\mathrm{c}=T_{\mathrm{c}-}$, and by $T_{\mathrm{c}+}$ which represents the memory of the lower $T_{\mathrm{c}\alpha}$. Importantly, that $T_{\mathrm{c}-}$ always increases, but $T_{\mathrm{c}+}$ decreases with $|W_{12}|$. By using single-band analogue, $Gi_d$ should raise with interband coupling, since the latter increases $T_{\mathrm{c}-}$, i.e. there should be monotone dependence $Gi_d(T_\mathrm{c})$. Here the two-gap nature comes into play via $\mathcal{F}_d(\kappa_-)$. By describing the whole two-band condensate, that channel governs fluctuations in a two-gap system around the phase transition point as well as near the instability of the weaker-superconductivity component taken as independent subsystem. This peculiarity of $\mathcal{F}_d(\kappa_-)$ involves $T_{\mathrm{c}+}$, the memory of lower $T_{\mathrm{c}\alpha}$, into considerations, as it would be additional critical point in the system. At that, by following one-band analogue, $Gi_d$ should reduce with interband coupling, since the latter suppresses $T_{\mathrm{c}+}$. Thus, there are two opposite tendencies associated with the temperatures $T_{\mathrm{c}\pm}$ which drive the behaviour of $Gi_d$. Their interplay becomes essential, when the former instability of the weaker-superconductivity component is located close to the phase transition point of the joint condensate and the memory of it is not completely erased by the interband interaction.

In the vicinity of phase transition point, the deviations from the mean-field predictions for the critical temperature, superfluid density, Josephson current, tunneling conductance (due to fluctuation induced suppression of density of states) etc. are defined by the value of $Gi_d$. It would be interesting to analyse these observables for different high-$T_\mathrm{c}$ materials keeping in mind that corrections to the Bardeen-Cooper-Schrieffer scenario from fluctuations can shed light on the presence of additional superconducting components as well as on the proportion between intrinsic superconductivities in the subsystems involved.

\section{Conclusions} We have demonstrated that the qualitatively new types of fluctuations must be taken into account in the thermodynamics of superconductivity if one involves several electron bands into the pairing mechanism. Together with gap-phase fluctuations, i.e. Goldstone and Leggett modes, there appear two distinct channels for gap-amplitude fluctuations in a two-band scenario. The distinctness is stemming from the corresponding correlation lengths which have remarkably different properties. The first one diverges at the phase transition point, $T_\mathrm{c}$, and refers also to the former criticality of the weaker-superconductivity constituent, smeared by interband interaction. Corresponding mode, representing actual as well as the former superconducting instabilities, dominates in the vicinity of phase transition point of the joint condensate. That involves effectively two critical temperatures into the evolution of the Ginzburg number. By manipulating the proximization between distinct electron subsystems, non-monotonic behaviour of the Ginzburg number with $T_\mathrm{c}$ can be produced, unlike single-band counterpart. At that, critical region can shrink up to $25\%$ of the value corresponding to the single-band limit. In such a way fluctuations reflect two-band nature of superconductivity. The second correlation length is always finite, but its temperature derivative has a jump at $T_\mathrm{c}$, defined uniquely by the interband coupling. The fluctuations related to this length scale should be taken into account as one explores two-band superconductivity outside the critical domain. 

\ack This study was supported by the European Union through
the European Regional Development Fund (Centre of Excellence "Mesosystems:
Theory and Applications", TK114), by the Estonian Science Foundation Grant
No 8991, by the Estonian Ministry of Education and Research through the Institutional Research Funding IUT2-27, and by the COST MP1201 NanoSC Action.

\section*{References}
\bibliography{2bandGi}
\end{document}